%% file: Cher2005_Bernardini.tex
\documentclass[12pt]{article}
\usepackage{epsfig,graphicx}
\usepackage{ch2005v1}

\begin{document}
\def\cer{\mbox{Cherenkov}}

\Title{Multi-Messenger Studies with AMANDA/IceCube: Observations and Strategies}
\label{BernardiniStart}
\author{Elisa Bernardini for the IceCube
Collaboration\\
\input{authors_EBenardini}}
\index{Bernardini, F.} 
\makeauthor
\abstracts{
Four years of AMANDA-II data have been searched for neutrinos from
point sources. No statistically significant excess of events has been
detected, neither integrated in the years 2000 to 2003, 
nor in the searches for occasional signals. 
An interesting coincidence of neutrinos  
with gamma-ray flares emerges when inspecting the
time of the events detected from the direction of the 
Blazar 1ES1959+650. The exceptional character of the 
gamma-ray observation provides a strong motivation for
consolidating similar search strategies with AMANDA and its successor
IceCube, as well as for multidisciplinary
investigations of this and other gamma-ray sources.
We report  the outcomes of the most recent survey of the northern sky
to search for neutrino point sources with AMANDA-II. We also discuss
possible viable collaborations between the gamma-ray and
the high energy neutrino observatories.}
\section{Introduction}
The primary goal of a neutrino telescope is the 
discovery of extraterrestrial neutrinos with high
energies. The research
field contributes to the increased understanding of the nature, the
origin and the propagation of cosmic rays. 
The detection of neutrinos is more challenging than 
that of cosmic rays and gamma-rays, due to the much
smaller cross section for neutrino interaction and therefore
the small detection probability. 
Neutrinos from point sources would  
provide an unambiguous signature of a
hadronic component in the flux of particles accelerated in the
astrophysical engines. Moreover, unlike protons and photons, 
neutrinos can propagate freely over cosmological distances. This, 
combined with the intrinsic complementary nature of this
observational window, motivates the search of cosmic neutrinos.

So far no excess of events ascribed to either a point-like or a
diffuse extraterrestrial flux of neutrinos has been 
observed~\cite{2000ps}-\cite{uhe}. Recently, four 
years of AMANDA-II data, collected between 2000 and 2003,
have been analyzed with improved reconstruction techniques and better
background rejection power compared to previous
publications.  
A large statistics sample of neutrinos
with high energies has been selected, allowing to search for point
sources with a sensitivity comparable to the observed gamma-ray fluxes
of Blazars, when in ``high state''~\cite{icrc4yr}. This indicates that neutrino
astrophysics is reaching discovery potential.

We report the results of a time-integrated
search of point sources of neutrinos, which provides
the most stringent flux upper limits currently 
available for the northern sky, and also the
first attempts to search for neutrino flares with AMANDA-II. 
A section is  dedicated to the observations of the Blazar
1ES1959+650. Finally, future perspectives of multidisciplinary
investigations of this and other objects are discussed.

\section{Operation principles of AMANDA and IceCube}
AMANDA-II is currently the largest operating neutrino telescope. Located
at the South Pole, the array comprises
677 optical modules to detect the $\cer$ photons from charged
particles in the ice shelf. Each module 
consists of a 8 inch diameter photo-multiplier, housed in a 
pressure-tight glass sphere. The optical modules 
are located at depths between 1.5 and 2 kilometers, in a
structure of 19 strings. The instrumented volume has a diameter of
about 200 meters~\cite{amanda2}. 
  
The ice overburden at the South Pole 
reduces the flux of cosmic muons to a 
rate less than 0.1 kHz, as measured with AMANDA-II. When muon neutrinos
with energies above a few tens of GeV undergo charged current interactions
in the ice surrounding or in the rock below the detector, muon tracks
emerge, which can be reconstructed based on the arrival time of the
$\cer$ photons at the  
optical modules. 
Due to the scattering of photons in ice, 
complex likelihood procedures are
necessary to achieve good angular resolution (between 2${^\circ}$ 
and 2.5${^\circ}$ for the typical track lengths in AMANDA-II). The muon
energy is estimated from the density of detected $\cer$ photons, with
an accuracy of 0.4 in the logarithm of the energy.

AMANDA is operating since 1996 (since 2000 as the full-scale  
AMANDA-II). In January 2005, one string of IceCube
was  installed and started
operation. IceCube will include 4800 optical modules (80 strings) in a
volume of 1 $\mbox{km}^3$. Apart from various technological
improvements compared to AMANDA, its geometry alone will ensure an
angular resolution comparable to the neutrino-muon scattering angle
(down to 0.5$^{\circ}$ at 5 TeV). IceCube is expected to achieve the
sensitivity to detect neutrinos from Active Galactic Nuclei and Gamma
Ray Bursts~\cite{learned}.

\section{Search for point sources in the northern sky}
Searches for astrophysical sources of neutrinos have to cope with the
background from interactions of cosmic rays with the Earth's
atmosphere. The dominating component stems 
from down-going muons and is suppressed with
angular cuts. This limits the searches for point sources essentially 
to the northern 
sky. A more uniform flux of neutrinos from meson decay
and a negligible fraction of mis-reconstructed cosmic muons remain,  
indistinguishable from cosmic neutrinos. These residual backgrounds 
are treated identically and their effect is evaluated statistically from the
density of the detected events as a function of declination, i.e. 
adopting a similar  approach as the ``off-source'' observations of gamma-ray
astronomy\footnote{Note that the geographic
location of the detector ensures a uniform and constant
exposition of portions of the sky at the same declination.}. 

A neutrino point source would manifest itself as a localized excess of events
on top of the  background. To ensure a high signal-to-noise ratio, 
the event reconstruction and selection are
optimized in a way to provide tracks with good angular resolution, in a
wide energy range. Details of the reconstruction algorithm can be
found in~\cite{recopaper}. Up-going events, induced by muon neutrinos, are
selected by imposing track quality requirements. 
\begin{figure}[!t]
\begin{center}
\hspace*{2mm}
\epsfig{file=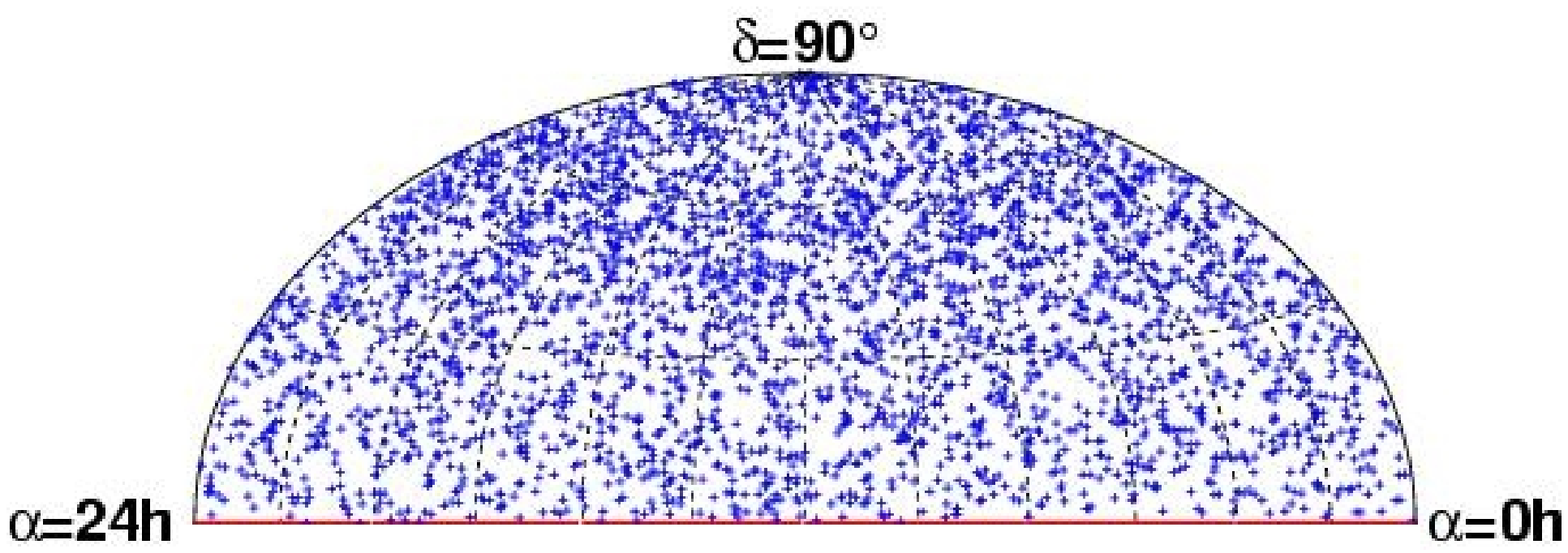,height=3cm}\epsfig{file=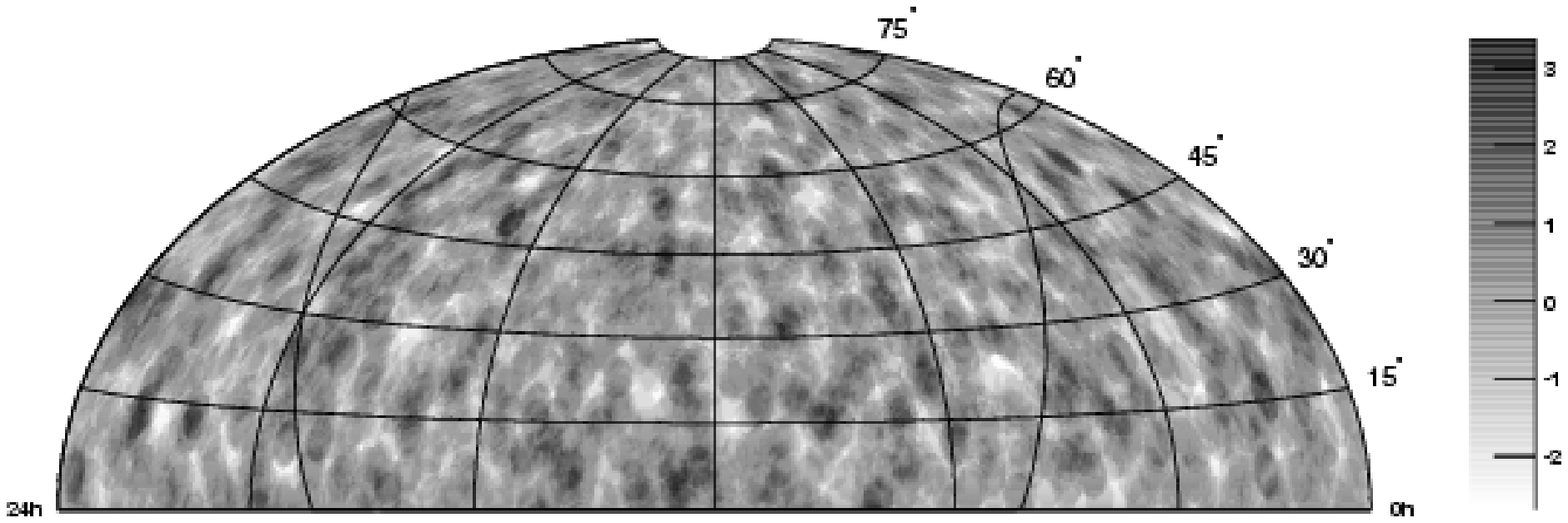,height=3cm}
\caption{Left: Sky map of neutrino events (3329 up-going tracks),
observed with AMANDA-II 
between the years 2000 and 2003. 
Right: Significance map of the search for clusters of events in the
northern sky based on the 3329 selected events.}
\label{fig:Bernardini-fig1}
\end{center}
\end{figure}
The total live-time considered is 807 days, after
data quality selection and rejection of the periods of detector
maintenance and station activities (between November and February). 
Event selection
criteria were optimized to achieve the best average flux upper limit
(sensitivity) for an assumed power-law signal energy spectrum 
with two extreme spectral indices: $\gamma$=2 and  
$\gamma$=3. Event cuts were optimized independently for different
declination bands. The search bin radius was a 
free parameter and varies between
2.25$^{\circ}$ and 3.75$^{\circ}$, depending on declination. 
The optimization procedure accounts for 
the effective live-time, which allows to loosen cuts
for dedicated investigations of sub-periods of data taking.

Our search for high energy cosmic neutrinos from known astrophysical
objects mostly focuses on  sources of high
energy gamma-rays. This choice is supported by the fact that
any object that accelerates charged hadrons to high energy is a likely
source of neutrinos: the hadrons will interact with other
nuclei or the ambient photon fields producing hadronic showers. In
these scenarios, high energy photons and neutrinos are expected to be
produced simultaneously.

\subsection{Time-integrated search}
\label{sec:Bernardini-time-integrated}
The sensitivity to point sources of neutrinos, with a
live-time of 807 days, is $6\cdot 10^{-8}$ GeVcm$^{-2}$s$^{-1}$ for a spectral
index $\gamma$=2, weakly dependent on declination.
A sample of 3329 up-going events was extracted,  
shown in Fig.~\ref{fig:Bernardini-fig1}-Left. 
Based on these events, we performed
a search for coincidences with the directions of a catalog of 33
selected objects, and also a full scan of the northern sky. In
both cases all observed excesses are compatible with the background
hypothesis. The significance of each observation was evaluated with
repeated and equivalent ``experiments'' performed on samples of events
obtained by randomizing the right ascension coordinates of the 3329
 neutrinos. This method allows a correct evaluation of the
trial factors in presence of multiple tested directions, and of the
correlations due to partial overlapping of the search
bins. 

Table~\ref{tab:Bernardini-tab1} 
summarizes the results of the test performed on
the catalog of 33 sources. The highest  excess
corresponds to the direction of the Crab Nebula (1.7~$\sigma$). 
The probability to
observe this or a larger excess due to a statistical fluctuation of
the background, in any of the 33 bins, is 64\%. The northern
sky was scanned with a system of highly overlapping bins, to maximize
the detection chance. The 
significance map is shown in Fig.~\ref{fig:Bernardini-fig1}-Right. 
The highest excess   
(3.4~$\sigma$) corresponds to a probability of a background
fluctuation of 92\%. The systematic uncertainty is under evaluation and
the flux upper limits will be reported in a forthcoming publication. The
preliminary results for the Blazars Markarian 421 and 1ES1959+650 are
0.7$\times$10$^{-8}$cm$^{-2}$s$^{-1}$ 
and 1.0$\times$10$^{-8}$cm$^{-2}$s$^{-1}$ respectively, for $\gamma$=2
 and integrated above 10 GeV. 
These results refer to 807 days of exposure. To compare them
to the observed high energy gamma-ray flares, for example from
Markarian 421~\cite{mrk421_0001}, it is necessary to introduce 
assumptions on the time the source was in a high state and
on the corresponding photon flux and 
spectral index. Considering X-ray light curves from~\cite{rxte}
we estimate an integral time of the order of 200 days of
``high activity'' of Markarian 421 between 2000 and 2003~\cite{icrctime}.
As gamma-ray 
flux and energy spectrum we assumed the results obtained for the
flares
 observed in 2000 and 2001, reported in~\cite{mrk421_0001}, and applied a
correction for the infra-red absorption according to~\cite{stecker}.
Including neutrino oscillation, we estimate a 
sensitivity to neutrinos from Markarian 421, for 200 days of
live-time, 
less than a factor 3 of the corresponding gamma-ray
flux, up to about 20 TeV.
\begin{table}[!h]
 \begin{center}
 \begin{footnotesize}
  \begin{tabular}{lcccc|lcccc}\hline\hline
      Candidate & $\delta$($^\circ$) & $\alpha$(h)    &
                $n_{\mathrm{obs}}$ & $n_{b}$          &
      Candidate & $\delta$($^\circ$) & $\alpha$(h)    &
                $n_{\mathrm{obs}}$ & $n_{b}$          
                 \\\hline
\multicolumn{10}{c}{ \emph{TeV Blazars} } \\
      Markarian 421  & 38.2 & 11.07 & 6 & 5.6 &
      1ES 2344+514   & 51.7 & 23.78 & 3 & 4.9 \\
      Markarian 501  & 39.8 & 16.90 & 5 & 5.0 &
      1ES 1959+650   & 65.1 & 20.00 & 5 & 3.7 \\ 
      1ES 1426+428   & 42.7 & 14.48 & 4 & 4.3 &
		     &&&&\\
\multicolumn{10}{c}{ \emph{GeV Blazars} } \\
      QSO 0528+134   & 13.4 &  5.52 & 4 & 5.0 &
      QSO 0219+428   & 42.9 &  2.38 & 4 & 4.3 \\
      QSO 0235+164   & 16.6 &  2.62 & 6 & 5.0 &
      QSO 0954+556   & 55.0 &  9.87 & 2 & 5.2 \\
      QSO 1611+343   & 34.4 & 16.24 & 5 & 5.2 &
      QSO 0716+714   & 71.3 &  7.36 & 1 & 3.3 \\  
      QSO 1633+382   & 38.2 & 16.59 & 4 & 5.6 &
		     &&&&\\
  \multicolumn{10}{c}{ \emph{Micro-quasars} } \\
      SS433          &  5.0 & 19.20 & 2 & 4.5 &
      Cygnus X3      & 41.0 & 20.54 & 6 & 5.0 \\
      GRS 1915+105   & 10.9 & 19.25 & 6 & 4.8 &
      XTE J1118+480  & 48.0 & 11.30 & 2 & 5.4 \\
      GRO J0422+32   & 32.9 &  4.36 & 5 & 5.1 &
      CI Cam         & 56.0 &  4.33 & 5 & 5.1 \\
      Cygnus X1      & 35.2 & 19.97 & 4 & 5.2 &
      LS I +61 303   & 61.2 &  2.68 & 3 & 3.7 \\
   \multicolumn{10}{c}{ \emph{SNR \& Pulsars} }\\
      SGR 1900+14    &  9.3 & 19.12 & 3 & 4.3 &
      Crab Nebula    & 22.0 &  5.58 &10 & 5.4 \\
      Geminga        & 17.9 &  6.57 & 3 & 5.2 & 
      Cassiopeia A   & 58.8 & 23.39 & 4 & 4.6 \\
  \multicolumn{10}{c}{ \emph{Miscellaneous} }\\
      3EG J0450+1105 & 11.4 &  4.82 & 6 & 4.7 &
      J2032+4131     & 41.5 & 20.54 & 6 & 5.3 \\
      M 87           & 12.4 & 12.51 & 4 & 4.9 &
      NGC 1275       & 41.5 &  3.33 & 4 & 5.3 \\
      UHE CR Doublet & 20.4 &  1.28 & 3 & 5.1 &
      UHE CR Triplet & 56.9 & 11.32 & 6 & 4.7 \\    
      AO 0535+26     & 26.3 &  5.65 & 5 & 5.0 &
      PSR J0205+6449 & 64.8 &  2.09 & 1 & 3.7 \\
      PSR 1951+32    & 32.9 & 19.88 & 2 & 5.1 &
		     &&&&\\
       \hline\hline
  \end{tabular}
 \end{footnotesize}
  \caption{\label{tab:Bernardini-tab1}
                Results from the search for neutrinos from selected
		objects, from the analysis of AMANDA-II data between
		2000 and 2003.
                $\delta$ is the declination in degrees, $\alpha$ the
		right ascension in hours,
                $n_{obs}$ is the number of observed events and $n_{b}$
                the expected 
                background.} 
\vspace*{-0.8cm}
 \end{center}
  \end{table}
\subsection{Search for neutrino flares}
The search of occasional flares of neutrinos in the sample of
selected up-going events is motivated by the high variability 
which characterizes the electromagnetic emission of many 
neutrino candidate sources. The flux upper limits derived from the
results reported in the previous section indicate that AMANDA-II has
achieved a sensitivity to neutrino fluxes which is comparable to the
observed high energy gamma-ray fluxes of Blazars in high states
(e.g. the flares of Markarian 501 in 1997~\cite{mrk501_97} and Markarian 421 in
2000/2001~\cite{mrk421_0001}). With the assumption that 
the (possible) neutrino emission would be characterized by a  flux
enhancement comparable to gamma-ray flares,
neutrino flares could be extracted from the sample of selected
events with a reasonable significance.
Under these considerations we developed a search for time-variable
neutrino signals from point sources following two different
approaches:
\begin{enumerate}
\item[a)] {\bf Search of clusters of neutrinos in coincidence with
known periods of enhanced electromagnetic emission of selected
objects}:
The objects and/or the periods of interest were chosen on the 
basis of a compilation of the  light curves reported at different
wavelengths. Due to the limited availability of high energy gamma-rays
observations, we referred to 
X-ray light curves for the two Blazars we considered
(Markarian 421 and 1ES1959+650~\cite{rxte}). For the third object, 
the Micro-quasar Cygnus X3, we instead used radio light
curves~\cite{ryle}. 
A proper re-optimization
of the neutrino event selection was performed, 
to account for shorter integrated
exposures compared to 807 days. The integrated periods-of-interests
were 141 days for Markarian 421, 283 days for 1ES1959+650 and 114 days for
Cygnus X3, based on threshold cuts on the X-ray/radio intensity curve.
\item[b)] {\bf Search of occasional neutrino flares 
from selected objects}: Twelve sources were considered, 
known to manifest a character of high
variability in the corresponding gamma-ray
emission. We considered four Blazars, four Micro-quasars and four
EGRET sources with exceptional variability in the MeV gamma-ray
emission. Neutrino flares have been searched for by comparing the observed
events with the time-dependent background, using sliding time-windows
fixed to 20 days duration for galactic objects and 40 days duration 
for extragalactic objects. This approach entails a higher trial factor
penalty than case a). 
As a merit, neutrino flares which are not accompanied by an
observed electromagnetic counterparts are not automatically excluded. 
This approach is 
also less dependent on models for the correlation between the neutrino
and the electromagnetic emission and not dependent on the
availability of multi-wavelength information. 
The test was performed on the sample of
3329 up-going events. The choice of both the window duration and
the test data sample was based on the outcome of
a dedicated Monte Carlo simulation. We considered the information
reported in Tab.~\ref{tab:Bernardini-tab1} and assumed hidden
neutrino flares with strengths compatible with the  
flux upper limits derived from Tab.~\ref{tab:Bernardini-tab1}. In other words,
we considered the maximum signal strength still compatible with the
background hypothesis at a 50\% confidence level. The search criteria were
optimized following a blind approach.
Events belonging to subsequent doublets
are assigned for simplicity to those clusters
showing the highest multiplicity or those 
occurring first, if having the same multiplicity. 
\end{enumerate}
The significance of each observation was evaluated in 
a similar way as described in the previous section. 
A proper treatment of the time variability of the
background was carried out. In all cases no statistically
significant excess was found. 

Event doublets were observed from the
directions of the Blazars 1ES1959+650, QSO 0235+164, the Micro-quasar
GRS 1915+105 and the EGRET sources 3EG J0450+1105, 3EG J1227+4302 
and 3EG J1928+1733, each with a background probability larger than 32\%.
No doublets were observed from the
directions of the Blazars Markarian 
421 and QSO 0528+134, of the Micro-quasars GRO J0422+32, Cygnus X1
and Cygnus X3, and the EGRET source 3EG J1828+1928.

\subsection{Neutrinos from the direction of the Blazar 1ES1959+650}
The Blazar 1ES1959+650 belongs to the catalog of the 33 tested
objects, reported in section~\ref{sec:Bernardini-time-integrated}. 
The search bin used
(2.25$^{\circ}$) contains between 65\% and 75\% 
of the Monte Carlo events passing the selection criteria. 
Five events have been
selected between 2000 and 2003, three out of five
within 66 days in the year 2002 (MJD 52394.0, 52429.0, 52460.3). 
This interval partly overlaps
with a period of exceptional activity of the source, monitored
by a multi-wavelength campaign (MJD between 52410 and
52500~\cite{multiwave}).  
A high energy gamma-ray
flare was observed without a corresponding counterpart in the X-ray
light curve. This event, referred to as an ``orphan flare'' is
generally considered as an indication of hadronic processes occurring
in the Blazar jet. Coincident high energy neutrinos are expected
in this case, although theoretical estimates of the expected fluxes
and of the discovery 
potential for AMANDA-II vary strongly~\cite{halzen,reimer}. 

\begin{figure}[!t]
\begin{center}
\hspace*{-0.5cm}\epsfig{file=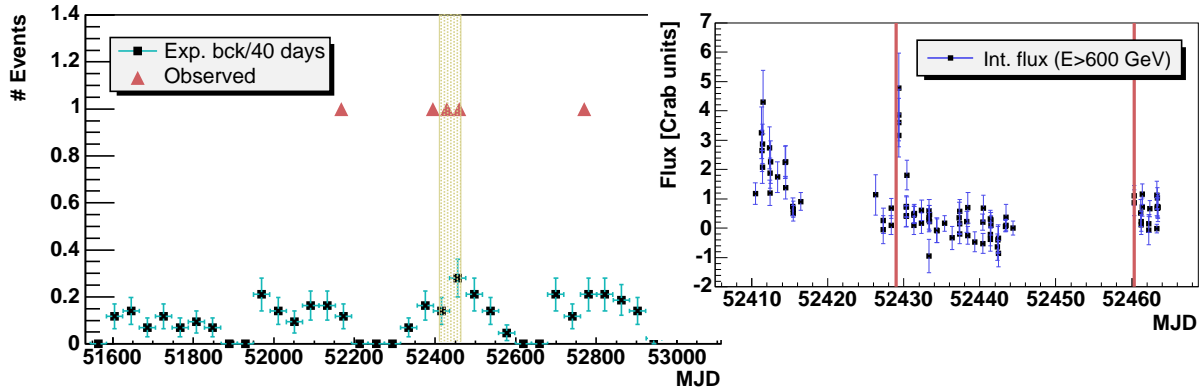,height=5.5cm}
\vspace{-1mm}
\caption{Left: AMANDA-II neutrino events observed within 2.25$^{\circ}$ from
the direction of the Blazar 1ES1959+650 (triangles). 
The crosses indicate the expected background per 40-day
bins (fluctuating due to variation in the effective
live-time). The hatched
area corresponds to the period of Whipple measurements
reported in~\cite{whipple}, shown at the Right. Superimposed (vertical
lines) are the
arrival times of the AMANDA-II events in the same period.}
\label{fig:Bernardini-fig3}
\end{center}
\end{figure}

One of the AMANDA-II neutrino events was recorded within
a few hours from the ``orphan flare''. The arrival times of the
observed neutrino events are plotted in Fig.~\ref{fig:Bernardini-fig3}-Left and
compared to the integrated background per 40-day windows. 
The significance of
the coincidence is low and it 
can not be easily quantified, due to the trial factors
arising from a-posteriori choices of the time windows to be used for
the statistical test. More
observations are necessary to shed light on the possible hadronic
nature of the particles emitted in the jet of this source.
\section{Viable perspectives for the multi-messenger approach}
Both the neutrino and the high energy gamma-ray
community aim to classify the nature of observed astrophysical objects
and to answer
the intriguing question whether the population of the accelerated
particles is purely electromagnetic, or mixed hadronic and 
electromagnetic.
In this context, the "orphan flares" detected from several
Active Galactic Nuclei are of particular interest.
An estimation of the frequency of these phenomena would
have strong implications on the understanding of the
origin of the observed cosmic
rays. The overlap of interests of the two communities extends even further
when other objects are
considered, like for example the unidentified EGRET sources.

Neutrino astrophysics is entering a new phase, with detectors
like AMANDA-II reaching a sensitivity region with discovery
potential and cubic-kilometer detectors being designed or~--~like IceCube --
already under construction and starting data-taking.
A  coordination of the efforts
between the gamma-ray and the neutrino communities
is going to be feasible and it should be of mutual
benefit in several ways.

First, the dramatically increased availability of 
data on high energy gamma-ray 
emission will allow a more qualified selection of
neutrino candidate sources and favorable periods 
then hitherto. A reduced set
of cases, with better founded expectations for the corresponding neutrino
emission, would limit the penalization from trial
factors and  enhance the discovery chance.

Second, neutrino observations might provide a
target-of-opportunity trigger to gamma-ray detectors.
With AMANDA-II we would look for
neutrino signals in coincidence with intense gamma-ray flares
which could be observed by small telescopes like HEGRA, as well as by the
third-generation
gamma-ray telescopes like CANGAROO, H.E.S.S., MAGIC, and VERITAS.
AMANDA-II is a continuously sensitive and large field-of-view telescope,
which allows the simultaneous and non-interrupted monitoring of all
sources located in the northern sky. Data is typically analyzed
off-line, following a blind procedure, i.e. event selections are
optimized in a way which avoids the introduction of statistical biases
(for example by adopting the ``off-source'' methodologies of gamma-ray
astronomy). 
An on-line reconstruction procedure has been
developed and its performance is being tested. "Neutrino triggers" based on
the on-line reconstruction could 
be provided to gamma-ray and
X-ray observatories within at most a few hours.
If, for a limited set of most promising objects,
a neutrino would be detected from 
the direction of one of these sources, gamma-ray telescopes
could promptly verify the corresponding level of activity.
Given a resolution for
the neutrino direction of about 2$^{\circ}$, most of the 
events will stem from atmospheric neutrinos.
Therefore a careful study of the expected ratio between true
and false alarms and the sustainable rate of
false alarms have to precede the implementation of such a ``hadronic trigger''. 

In case of sources which are already included in the
scientific program of the involved experiments, neutrino-based
target-of-opportunity measurements may
entail no extra observation time. Table~\ref{tab:Bernardini-tab1}
provides an indication of the trigger rate of neutrino events from the
direction of selected objects, with the cut strength adopted for this
analysis. In particular the chosen sky bins contain a
fraction of the signal Monte Carlo
events, passing the same selection, which varies between
about 60\% up to about 85\%, according to the source declination and
the assumed spectral index.

In conclusion, we encourage the long-term and unbiased monitoring
at different wavelengths of
those neutrino candidate sources  which show an
evident character of variability in the high energy gamma-ray
emission (Blazars in particular). 
We also encourage the establishment of working groups
to further develop the multi-messenger approach, i.e. to involve neutrino
observations within the already effective multi-wavelength
campaigns, and, in general, multidisciplinary investigations of
objects like the Blazar 1ES1959+650 and similar. 

\section*{Acknowledgments}
\begin{footnotesize}
We acknowledge the support of the following agencies: National
Science Foundation--Office of Polar Programs, National Science
Foundation--Physics Division, University of Wisconsin Alumni Research
Foundation, Department of Energy, and National Energy Research
Scientific Computing Center (supported by the Office of Energy
Research of the Department of Energy), UC-Irvine AENEAS Supercomputer
Facility, USA; Swedish Research Council, Swedish Polar Research
Secretariat, and Knut and Alice Wallenberg Foundation, Sweden; German
Ministry for Education and Research, Deutsche Forschungsgemeinschaft
(DFG), Germany; Fund for Scientific Research (FNRS-FWO), Flanders
Institute to encourage scientific and technological research in
industry (IWT), and Belgian Federal Office for Scientific, Technical
and Cultural affairs (OSTC).
\end{footnotesize}

\label{BernardiniEnd}
 
\end{document}

%% file: authors_EBenardini.tex
A.~Achterberg$^{t}$, 
M.~Ackermann$^{d}$, 
J.~Ahrens$^{k}$, 
D.W.~Atlee$^{h}$, 
J.N.~Bahcall$^{u}$, 
X.~Bai$^{a}$, 
B.~Baret$^{s}$, 
M.~Bartelt$^{n}$, 
R.~Bay$^{i}$, 
S.W.~Barwick$^{j}$
% ...
\\
${(a)}$ Bartol Research Institute, University of Delaware, Newark, DE 19716 USA\\
${(b)}$ Department of Physics, University of Wuppertal, D-42119 Wuppertal, Germany\\
${(c)}$ Universit\'e Libre de Bruxelles, Science Faculty CP230, B-1050 Brussels, Belgium\\
${(d)}$ DESY, D-15735, Zeuthen, Germany\\
${(e)}$ Blackett Laboratory, Imperial College, London SW7 2BW, UK\\
${(f)}$ Dept. of Technology, Kalmar University, S-39182 Kalmar, Sweden\\
${(g)}$ Lawrence Berkeley National Laboratory, Berkeley, CA 94720, USA\\
${(h)}$ Dept. of Physics, Pennsylvania State University, University Park, PA 16802, USA\\
${(i)}$ Dept. of Physics, University of California, Berkeley, CA 94720, USA\\
${(j)}$ Dept. of Physics and Astronomy, University of California, Irvine, CA 92697, USA\\
${(k)}$ Institute of Physics, University of Mainz, Staudinger Weg 7,~D-55099~Mainz,~Germany\\
${(l)}$ Dept. of Physics, University of Maryland, College Park, MD 20742, USA\\
${(m)}$ University of Mons-Hainaut, 7000 Mons, Belgium\\
${(n)}$ Dept. of Physics, Universit\"at Dortmund, D-44221 Dortmund, Germany\\
${(o)}$ Dept. of Physics, University of Wisconsin, Madison, WI 53706, USA\\
${(p)}$ Dept. of Physics, University of Wisconsin, River Falls, WI 54022, USA\\
${(q)}$ Division of High Energy Physics, Uppsala University, S-75121 Uppsala, Sweden\\
${(r)}$ Dept. of Physics, Stockholm University, SE-10691 Stockholm, Sweden\\
${(s)}$ Vrije Universiteit Brussel, Dienst ELEM, B-1050 Brussels, Belgium\\
${(t)}$ Dept. of Physics and Astronomy, Utrecht University, NL-3584 CC Utrecht, NL\\
${(u)}$ Institute for Advanced Study, Princeton, NJ 08540, USA\\
${(v)}$ Dept. of Physics and Astronomy, University of Kansas, Lawrence, KS 66045, USA\\
${(w)}$ Dept. of Physics, Southern University, Baton Rouge, LA 70813, USA\\
${(x)}$ Dept. of Astronomy, University of Wisconsin, Madison, WI 53706, USA\\
${(y)}$ Dept. of Physics, Chiba University, Chiba 263-8522 Japan\\
${(z)}$ CTSPS, Clark-Atlanta University, Atlanta, GA 30314, USA\\
${(aa)}$ Dept. of Astronomy and Astrophysics, Pennsylvania State
University, University Park, PA 16802, USA\\
${(ab)}$ Institut f\"ur Physik, Humboldt Universit\"at zu Berlin,
D-12489 Berlin, Germany